Dario Rodighiero, Lins Derry, Douglas Duhaime, Jordan Kruguer,
Maximilian C. Mueller, Christopher Pietsch, Jeffrey T. Schnapp, Jeff Steward & metaLAB

# Surprise machines

## Revealing Harvard Art Museums' image collection

**Surprise Machines is a project of experimental museology that sets out to visualize the entire image collection of the Harvard Art Museums, with a view to opening up unexpected vistas on more than 200,000 objects usually inaccessible to visitors. The project is part of the exhibition organized by metaLAB (at) Harvard entitled Curatorial A(i)gents and explores the limits of artificial intelligence to display a large set of images and create surprise among visitors. To achieve this feeling of surprise, a choreographic interface was designed to connect the audience's movement with several unique views of the collection.**

## 1. Introduction

Although "the humanities so far has focused on literary texts, historical text records, and spatial data," as stated by Lev Manovich in *Cultural Analytics* (Manovich, 2020, p. 10), the recent advancements in artificial intelligence are driving more attention to other media. For example, disciplines such as digital humanities now embrace more diverse types of corpora (Champion, 2016). Yet this shift of attention is also visible in museums, which recently took a step forward by establishing the field of experimental museology (Kenderdine et al., 2021).

This article illustrates the visualization of an extensive image collection through digital means. Following a growing interest in the digital mapping of images—proved by the various scientific articles published on the subject (Bludau et al., 2021; Crockett, 2019; Seguin, 2018), Ph.D. theses (Kräutli, 2016; Vane, 2019), software (American Museum of Natural History, 2020/2022; Diagne et al., 2018; Pietsch, 2018/2022), and presentations (Benedetti, 2022; Klinke, 2021)—this text describes an interdisciplinary experiment at the intersection of information design, experimental museology, and cultural analytics.

Surprise Machines is a data visualization that maps more than 200,000 digital images of the Harvard Art Museums (HAM) and a digital installation for museum visitors to understand the collection's vastness. Part of a temporary exhibition organized by metaLAB (at) Harvard and entitled Curatorial A(i)gents, Surprise Machines is enriched by a choreographic interface that allows visitors to interact with the visualization through a camera capturing body gestures. The project is unique for its interdisciplinarity, looking at the prestigious collection of Harvard University through cutting-edge techniques of AI.

The text structure illustrates to the reader 2. the museum and its collection, 3. the curatorship philosophy behind the exhibition, 4. the technical solutions adopted for the visualization, 5. the software for touchless interaction, and 6. how the team coped with technical drawbacks.

## 2. Harvard Art Museums

The story of the Harvard Art Museums starts in 1891 with a generous donation by Mrs. Elizabeth Fogg to establish an art museum in memory of her husband, William Hayes. A few years later, the architect Richard Morris Hunt built the Fogg Art Museum at the heart of Harvard University, which witnessed its opening to the public in 1895. On its hundredth anniversary, Director James Cuno describes the rapid growth of art collections by reporting that "more than 450 new art museums were built and more than 100 existing museums added over ten million square feet of new space" (Cuno, 1996). Cuno also writes that Harvard University was not an exception, raising new spaces for the Arthur M. Sackler Museum and the Busch-Reisinger Museum to host an extensive collection of 150,000 works. This collection is now part of the Harvard Art Museums, which integrated the three museums into a unique institution in 1983.

Since the publication of James Cuno's volume in 1996, Harvard Art Museums has exceeded the number of 200,000 works. Although the several extensions of the landmark Fogg building, the collection's continuous growth led to a complete restoration, completed in 2014 by the architect Renzo Piano. After the demolition of post-1925 expansions, an entirely new section facing Prescott Street integrated the original building of Quincy Street. In addition, a new glass rooftop structure brought natural light into the galleries, the laboratories, and the original courtyard in Travertine marble (Jodidio, 2014).

Around the same time, the digital turn entered the world of museums favoring archival work but also a further expansion of archival capacity. Haidy Geismar describes this transformation through the term *contact zone*, "where old museum collections and new technologies come together, tracing the translation and extension of collections from card catalogues, storerooms and display cases into digital websites, imaging platforms and collection management systems" (Geismar, 2018). In this twist, the Harvard Art Museums created an information system that counts almost 240,000 digital objects, composed of high-resolution images and metadata containing titles, descriptions, attributions, dates, classifications, credits, authors, subjects, media, dimensions, and provenances. All these objects are publicly accessible online (Harvard Art Museums, 2012) through the International Image Interoperability Framework (IIIF), which provides a web protocol for image and metadata delivery (Snydman et al., 2015).

In the Harvard Art Museums, the contact zone between physical and digital collections materializes in the multimedia space called Lightbox Gallery. Situated on the top floor, the gallery receives its name from the diffused light that comes from the glass rooftop structure. Conceived by metaLAB and HAM to bring visitors digitally in contact with the collection, the gallery features a large wall screen composed of nine interconnected monitors. As stated on the HAM website, "The Lightbox Gallery is a venue for digital experimentation—a space for projects that respond to museums' collections through new media and emerging technologies." This definition perfectly fits the domain of experimental museology in which immersive visualization and cultural data are crucial today (Kenderdine et al., 2021).

## 3. Can machines curate?

The Curatorial A(i)gents exhibition took place at the Harvard Art Museums in spring 2022; the poster in Figure 1 shows the richness of the exhibition's program. The exhibition presented a series of machine-learning digital installations by metaLAB (at) Harvard, a laboratory described as "an experimental platform that seeks to model new forms of cultural communication, creative practice, and scientific knowledge production" (Birkle & Däwes, 2019, p. 112; metaLAB, 2022). The curatorial philosophy of the exhibition appeared in a pamphlet edited by Mike Maizels and designed by Chelsea Qiu, which collects short essays by authors and guests (Maizels & Qiu, 2020). In his article entitled *iQueries*, the metaLAB founder Jeffrey Schnapp discloses the meaning of Curatorial A(i)gents. He alludes to the curator as the figure "who serves as a relay between museum collections and museum programming," in combination with the neologism a(i)gent, "as the mark of an encroachment on a terrain […] now carried out via the forms of […] 'artificial intelligence'" (Maizels & Qiu, 2020). Then Schnapp gets to the heart of the matter by formulating the question: can machines replace human curators? The provocative approach usually employed to point out digital inequalities (DiMaggio & Hargittai, 2001) and technology biases (Crawford & Paglen, 2019) is here shifted to investigate innovative curatorial practices through AI techniques.

Surprise Machines owes its name to Alan Turing's *imitation game*, described in the pioneering article "Computing Machines and Intelligence" (Turing, 1950). Can machines think? Turing answered using an experiment in which an examiner tries to distinguish between humans and machines while communicating via a typewriter; the machine is understood as engaged in thought when the responses appear indistinguishable. In the article, Turing develops an argument about AI from multiple perspectives, one of which was inspired by the mathematician Ada Lovelace (McCully, 2019). When Lovelace argued that machines are incapable of thought due to their inability to *take us by surprise*, Turing countered by stating that machines are a frequent source of astonishment due to their unpredictable behavior— thereby generating surprises.

**Figure 1.** This image shows the complete program of Curatorial A(i)gents, which alternates between weekly digital installations and online discussions with the authors

Surprise Machines thus aims to *surprise* visitors by showing them about 213,000 digitized images from the Harvard Art Museums' collection (a number that is slightly smaller than the entire collection as not all the objects went through digitalization). On the one hand, visitors are amazed by the size of the collection; on the other hand, they are astonished by the small number of objects showcased in the HAM building (a quick estimate from the website proves that less than one percent of the objects have been made visible to visitors over the years). In addition, an experienced eye may be surprised by more specific cues, such as the considerable number of Bauhaus drawings or the unique assortments of powdered pigments.

## 4. How to map 200,000 images

From the very first moment, there was no doubt that Curatorial A(i)gents needed one project to reveal the entirety of the HAM collection. As Norman Foster's glass-domed Reichstag plays with the analogy of political transparency (Foster, 2011), we imagined Renzo Piano's glass rooftop structure as the way to shed light on the HAM collection by using data visualization.

Data visualization is a computational and design practice aimed at revealing insights by translating tabular data into visual information. In a recent interview, Manuel Lima articulates the translation process through Nathan Shedroff's diagram (Rodighiero, 2021b). Yet this process is more technically and intellectually challenging when the dataset's richness has to be represented in its integrity, introducing a high level of *complexity* in the visual outcome. Lima is probably among the first authors to tackle this subject by creating an archive of network visualizations. Ten years after the publication of Visual Complexity (Lima, 2011), the landscape of network visualization has changed considerably thanks to scientists such as Albert-László Barabási (Barabási et al., 2020) who has played a central role in retracing the epistemic trajectory of complexity (Weaver, 1948). Among new algorithms, the replacement of force-oriented layouts (Bostock et al., 2011; Jacomy et al., 2014) with dimensionality reduction (Maaten & Hinton, 2008; McInnes et al., 2018) is key to handling larger relational datasets. The complexity of mapping extensive collections of images stems from the problem of computing and interacting with a considerable number of elements simultaneously.

For Surprise Machines, a local dataset of images was created by downloading the IIIF manifestos and their files. This operation was possible through the Application Programming Interface served by the Harvard Art Museums (Steward, 2015/2021), which provided about 213,000 JSON and JPG files by taking up 50 GB of disk space. Unfortunately, the first attempt to compute lexical metrics from text annotations was proved unsuccessful for two reasons: on the one hand, the high number of elements undermined the computation time of the force-directed graph layout (Bostock et al., 2011); on the other hand, many objects poorly annotated were unconnected to the network visualization. Although this method previously led to promising results (Moon & Rodighiero, 2020; Rodighiero, 2021a; Rodighiero et al., 2021, 2022), this was not the case. The solution came through a computational study of Aby Warburg's Atlas Mnemosyne (Impett & Moretti, 2017), which led to a collaboration with the Yale University Digital Humanities Laboratory (DHLAB).

The Yale University DHLAB supports scholars interested in digital methods. Among the digital tools provided by the library unit, the PixPlot tool was a significant breakthrough for Surprise Machines. PixPlot is a software composed of two parts, one in Python for processing and one in JavaScript for displaying (Duhaime, 2017/2021). Python code initially processes a set of bitmap images with Inception Convolutional Neural Network (O'Shea & Nash, 2015) trained on ImageNet 2012 (Russakovsky et al., 2015). This code transforms JPG files into vectors consisting of numbers. These numbers are successively computed using a technique of dimensionality reduction called UMAP (McInnes et al., 2018) to create a metric of image similarity. This algorithm returns a list of two-dimensional coordinates through a process that may seem opaque to many but that can be explained (Karjus et al., 2022). Finally, JavaScript code creates a web-based, zoomable interface that employs WebGL to speed up the rendering (Danchilla, 2012).

The results coming from PixPlot for visualizing the HAM collection were satisfying: the network visualization was balanced and evenly distributed on the screen; more than 200,000 images were spaced out in clusters by visual similarity (see Figure 2). The effect looked like

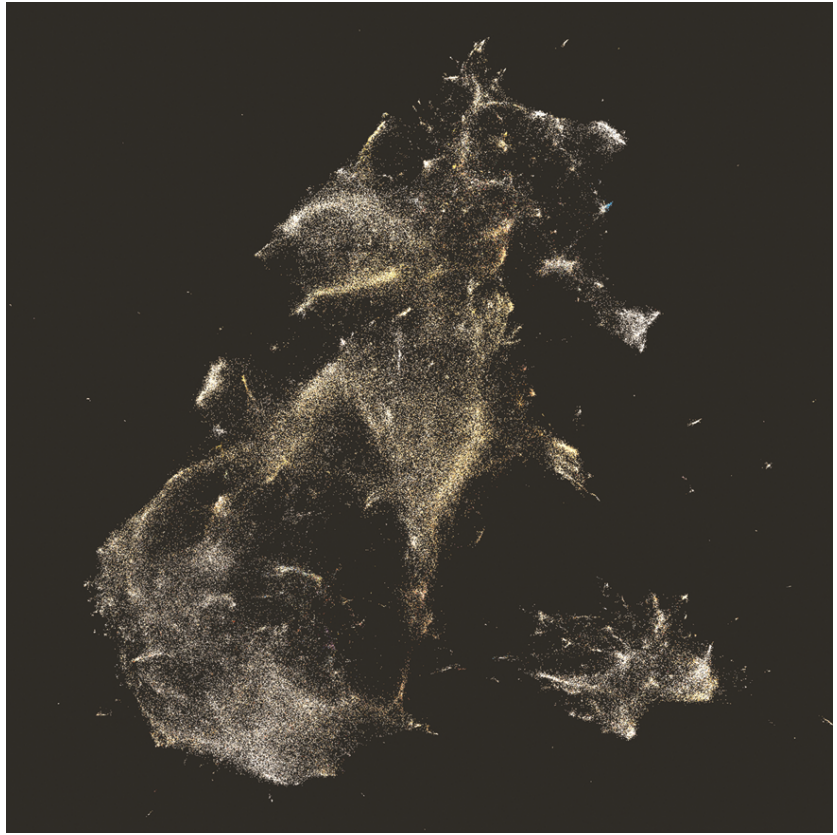

**Figure 2.** This network visualization presents Harvard Art Museums' collection through PixPlot, an open-access software created by the Yale University Digital Humanities Lab. The nebula shows the images in the two-dimensional space by visual similarity

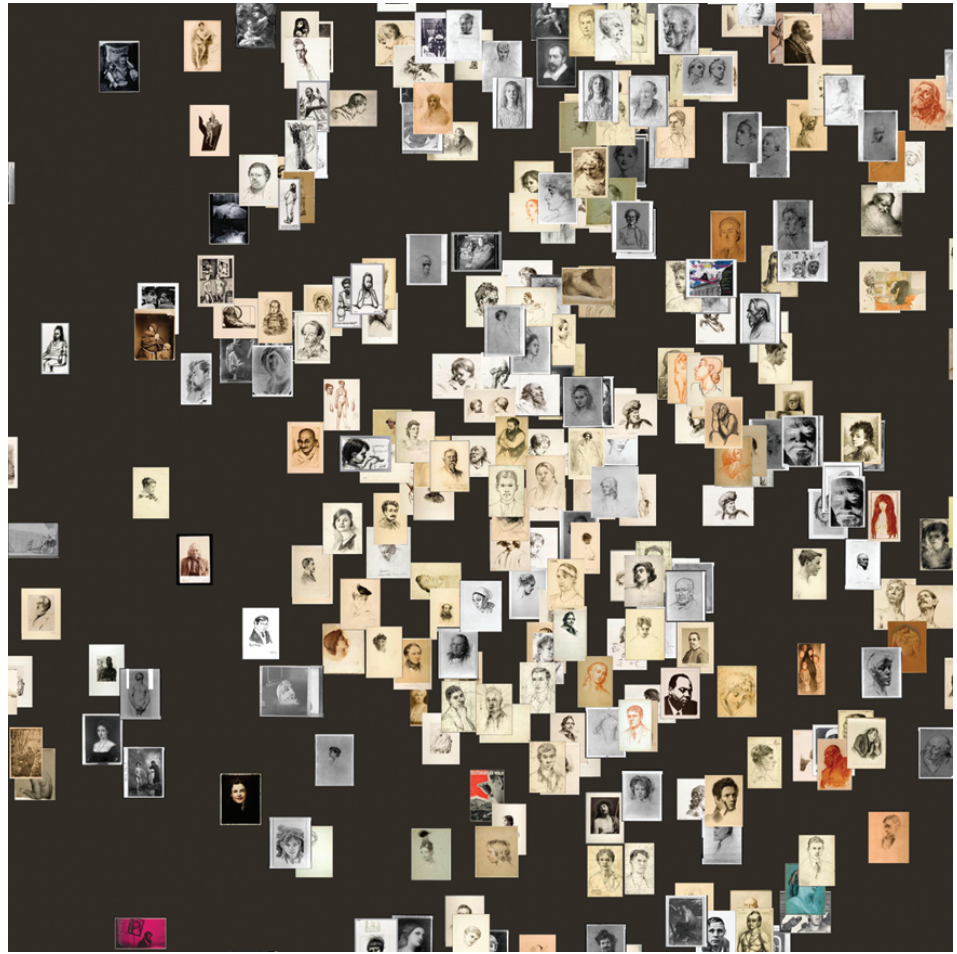

**Figure 3.** By using the web-based interactive interface, the visitor may zoom into specific areas of the visualization to reveal image clusters of Harvard Art Museums' collection. In this area showing hand-made portraits, Mahatma Gandhi and Albert Einstein are visible at the top left of the center

a nebula of points, ready to be explored by zooming through the web interface. Figure 3 enlarges one area of Surprise Machines, revealing a cluster of handmade portraits in which Mahatma Gandhi and Albert Einstein can be recognized just to the top left of the center.

Using a concept developed by Lev Manovich, the fascination of visualizing extensive collections is purely *sublime* (Manovich, 2008), but zooming into details brings information. The attentive use of the interface enables the discovery of specific clusters such as the already-mentioned Bauhaus drawings or powdered pigments. The designers who created the visualization leave room for viewers whose performative insights change from person to person (Drucker, 2013).

## 5. Designing a post-pandemic, choreographic interface

Curatorial A(i)gents was slated to open in 2020 but was postponed by two years due to the pandemic. The 11 projects comprising the show were all screen-based, with about half requiring conventional mouse and keyboard interactions like clicking and scrolling. Uncertain about the health protocols in place when the show opened, metaLAB proactively sought a solution for touchless interfacing with the projects. The assembled team saw this as an opportunity to research and develop a "choreographic interface" that would enable museum visitors to use a gestural vocabulary for exploring the projects.

Conceptually, the choreographic interface is a "human-computer interface that increases the kinetic and spatial interactivity between humans and computers through integrating *choreographic thinking* into the design process" (Derry et al., 2022). Compositional models pertaining to movement, space, and time were used to develop a full-torso gestural vocabulary specific to the projects' interaction needs. These included tracking, selecting, dragging, zooming, scrolling, right/left advancement, hand-switching, and browser refresh behaviors (see Figure 4). To interpret the choreographed gestures, we used open-source machine vision and machine learning tools such as OpenCV, TeachableMachine, and MediaPipe. Significant iterations were made in response to these technologies' constraints; these tools are limited to only processing static positions that are extremely distinct from one another. This led us to choreograph torso gestures that are highly geometric and dissimilar, and that sculpt the negative space about the body to make the limbs always visible. Over many iterations, we developed a vocabulary that balances choreographic interest and computational legibility. Much of the prototyping affected Surprise Machines because it included the most challenging pack of interactions. For example, managing the choreography of hand tracking, selecting, dragging, and switching without one method being misinterpreted as another is difficult since all these gestures center on the hand.

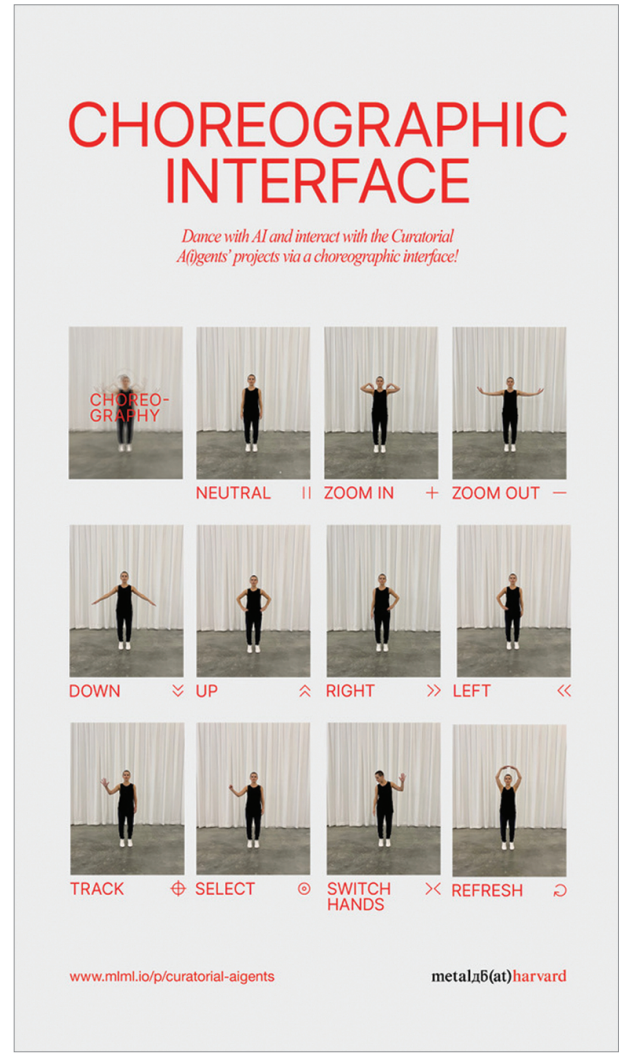

**Figure 4.** The choreographic interface invites visitors to interact with the Curatorial A(i)gents projects using a gestural vocabulary. This poster by Pablo Castillo shows the gestures for neutral, zoom in and out, scroll down and up, advance right and left, track, select (also used for dragging), switch hands, and refresh

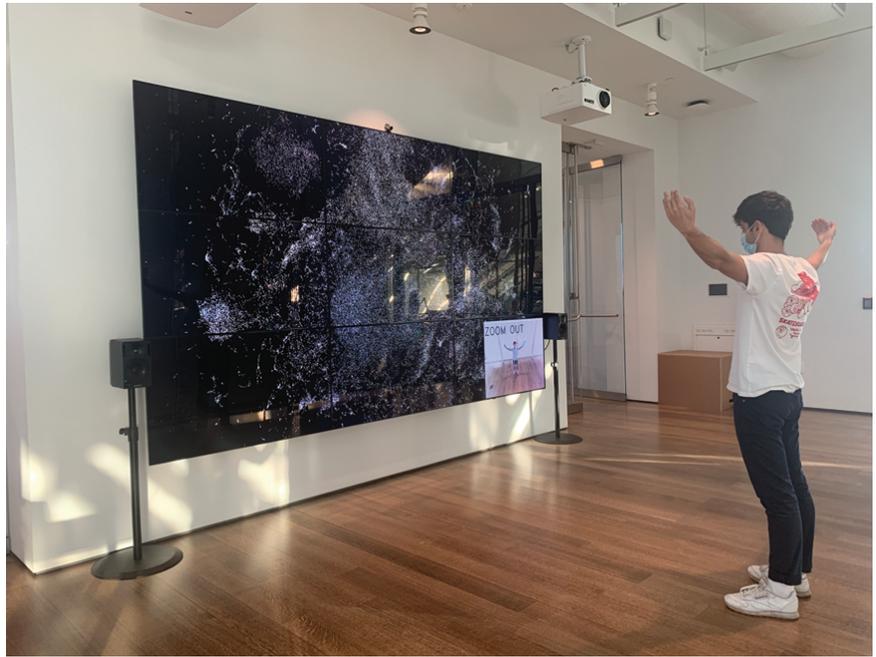

**Figure 5.** Surprise Machines test at the Harvard Art Museum's Lightbox Gallery showing the choreographic interface in action. The figure also clearly illustrates the Trails' interface based on a limited set of image previews, distinguishable by the white background

Surprise Machines requires simultaneous tracking and identification of the hand position and torso gestures. Our early solutions were computationally heavy and practically non-performing, especially when running the earlier version of Surprise Machines that used PixPlot. We eventually achieved faster performance by using MediaPipe's Holistic model for pose estimation and a logistic regression model for the classifier; Surprise Machines' shift to Trails, which is a lighter application, also boosted performance.

Prototyping for Surprise Machines also helped us think through the intuitiveness and relationships of the gestures. For example, we settled on hands-to-shoulders for zooming in; and arms outstretched in a "T" shape for zooming out to echo the sensation of bringing the visualization *into* the body and *away* from the body while interacting. Finding it awkward to move one's dominant hand across the body to the upper corner on the opposite side of the screen, we implemented a simple method for switching dominant hands by looking at the passive hand. Wishing to provide visitors with a performative experience while using the choreographic interface, we added sonification as our last embellishment.

The sonification applies and modulates subtle audio textures to the gestures, providing a dynamic score for the dance and reinforcing to the interface actions. The body's position and acceleration drive changes to the sound in service of marrying physical and digital movement. While the choreographic interface includes visual feedback on the screen (see Figure 5), aural feedback frees the visitor's visual attention allowing them to concentrate more fully on the projects.

## 6. A substantial drawback

Testing was a delicate phase of the project, especially considering that Surprise Machines was developed by a team whose members were based in different parts of the world. Thankfully, some of us were at the Harvard Art Museums when lockdown measures against COVID-19 were less strict, which made testing possible.

When the first test for data visualization took place, the team had mixed feelings. Although the visual was incredibly captivating, the hardware did not support the computation load because of the high number of images shown simultaneously. As a result, another software called Trails—conceived by DHLAB for large screens— was adopted to create a lighter version of Surprise Machines. Even though the back ends of PixPlot and Trails were similar, the front end's computational load was reduced by replacing the images with distributed samples (see Figures 6 and 7). The new interface was less sophisticated but way faster and more reactive, a relevant feature considering that the Lightbox Gallery's hardware also managed the choreographic interface's load.

When part of the team was working on the visualization between the U.S. and Europe, another

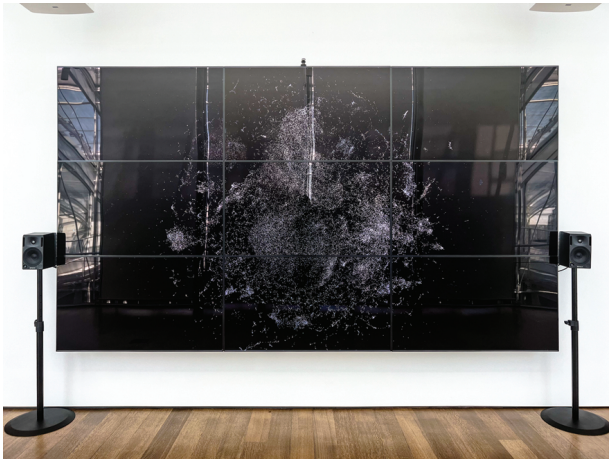
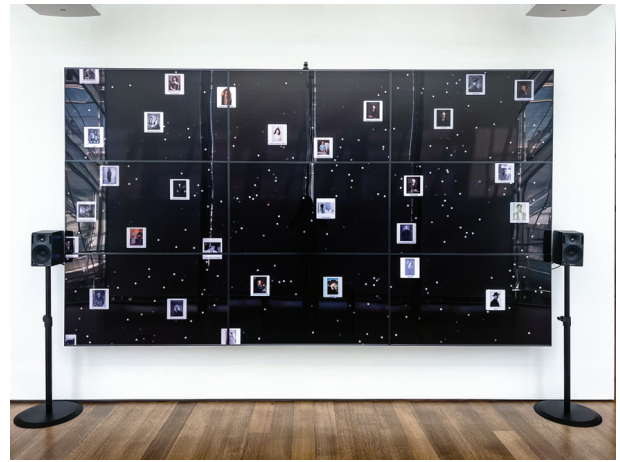

**Figures 6 and 7.** These are two photographs of Surprise Machines, taken by Sarah Newman during the exhibition: The first one shows the nebula of more than 200,000 images from the HAM's collection, the second one shows a frozen screenshot after zooming in. The latter represents the sampling used to lighten the interface with the remaining images in the background as circles part was dispersed across U.S. states designing the choreographic interface. While fortunate to run a few remote and in-person tests at HAM, problem-solving for the Lightbox Gallery's system was challenging as the team was developing Surprise Machines on different hardware setups. The operating system, hardware age, and hardware setup that included the camera, speakers, and nine interconnected monitors were significantly different. Although a choreographic interface for Surprise Machines was fully functional, it was not robust enough to be left installed for the exhibition's 11-week duration. Since public health protocols did not inhibit the use of shared devices in spring 2022, the final installation featured an air mouse to interact with the projects. Nonetheless, we look forward to including the choreographic interface for future Surprise Machines renditions when the tech driving the system can be managed more closely.

The Lightbox Gallery finally hosts Surprise Machines during the week of May 17–22, 2022. An air mouse was at visitors' disposal to move the cursor as if it were running on a standard personal computer, featuring dragging and dropping as well as zooming in and out. While most experienced visitors interacted with the digital installation without help, some had difficulty due to limited literacy, as previously discussed in a scientific article by Katy Börner and her colleagues (Börner et al., 2016).

Even with the increase in data literacy, new technologies still struggle to enter museum spaces. The Harvard Art Museums transformed a part of the building Piano envisioned as a meeting room into a laboratory dedicated to experimental museology. However, Curatorial A(i)gents and Living by Protocol, the two back-to-back exhibitions curated by metaLAB, were the penultimate programming in the Lightbox Gallery. Conceived in July 2012 and opened in November 2014, the Lightbox Gallery was converted into a quiet lounge for visitors in summer/fall 2022. The reason for this second conversion was twofold: on the one hand, the Lightbox Gallery was seen as a space with a cycle predestined to end; on the other hand, there was a lack of support within the museum, whose funding for staffing and maintenance ceased in 2018. The lack of funding was the primary reason for the closure of the director's office in July 2021. Although the detachment of the Lightbox Gallery from the main exhibition stories gave great freedom of planning, the truth is that Renzo Piano's restoration did not consider the presence of digital devices within the museum at all, and the result is a building extraordinarily crafted but classically designed. It means that the Harvard Art Museums may still live without any digital faculty but also that the digital turn has still to reach its full potential in museums. The experience Lightbox Gallery leaves is an increased understanding of experimental museology for the staff and the authors who exhibited in the gallery and new knowledge for designing future spaces.

## 7. Conclusions

Beyond the technical merits, Surprise Machines is a collective project developed in a stimulating research community above all. Resting on a solid curatorship philosophy, the project embraces a collaborative spirit that finds its most prominent expression in the choreographic interface. Although the latter was not part of the exhibition due to technical limitations, it is valuable to look at failures from a future perspective. Failures are indeed necessary sources of knowledge, as was the case with the closure of the Lightbox Gallery after a life cycle of about ten years. When thinking about a project such as Surprise Machines, it is necessary not to isolate the data visualization from the context. Authors, organizations, buildings, and technologies are among the human and non-human actors that provide extraordinary richness and unpredictability to the outcome. In a research environment where works become increasingly sectoral and specific, it is essential to consider *multidisciplinarity* as one of the most noteworthy qualities of a laboratory (Manzini, 2016), and *irreductionism* as an inexhaustible source of creativity and inspiration (Latour, 1988).

*About the authors*

**Dario** is an Assistant Professor of Sciences and Technology Studies at the University of Groningen, serving the multidisciplinary faculty Campus Fryslân. His research is rooted in knowledge design, critical data studies, and digital humanities; focusing on data and architectures in complex systems, cultural heritage, and open science.

Email: d.rodighiero@rug.nl

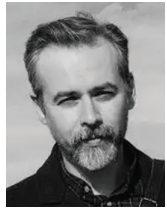

**Lins** is a Principal at metaLAB (at) Harvard working in the domains of human-computer interaction, data communication, and choreography. At metaLAB, she researches how choreographic interfaces and data embodiment can be developed toward functional and aesthetic ends. Prior to this, she danced professionally in New York, San Francisco, and Montréal.

Email: lins@metalab.harvard.edu

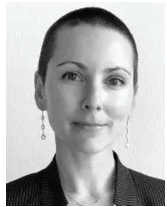

**Douglas** is an experienced software engineer with a demonstrated history of architecting and implementing solutions to complex business problems across a variety of market sectors and technology platforms. He is currently employed at Y Combinator and was previously the lead software engineer at the Yale University Digital Humanities Laboratory.

Email: douglas.duhaime@gmail.com

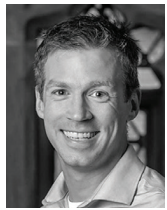

**Jordan** graduated from Harvard College where he studied applied math. In his studies, work, and research Jordan explores problems at the intersections of physics, engineering, machine learning, materials, and design. With metaLAB Jordan has contributed to the design and development of the choreographic interface for the Curatorial A(i)gents project.

Email: jordan@metalab.harvard.edu

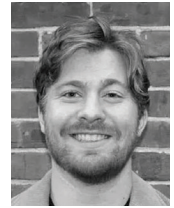

**Maximilian** is an artist and designer with a scientific foundation interested in ambient communication and reactive media. He is a graduate student in Design Engineering at the Harvard Graduate School of Design and Harvard John A. Paulson School of Engineering and Applied Sciences.

Email: maximilianmueller@mde.harvard.edu

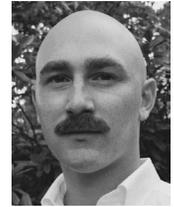

**Christopher** is an interaction designer who crafts digital experiences. His passion for interfaces goes beyond the digital layer, trying to connect the digital world with the physical one. As a freelancer, he explores novel types of visualization metaphors. He is affiliated with the Urban Complexity Lab at the University of Applied Sciences Potsdam.

Email: cpietsch@gmail.com

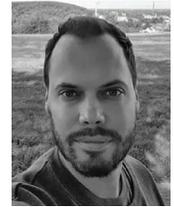

**Jeffrey** is a cultural historian, designer, curator, and technologist with research interests extending from antiquity to the present. At Harvard he occupies the Carl A. Pescosolido Chair in Romance and Comparative Literatures, serving as a faculty member of the Architecture department and as one of the faculty co-director of the Berkman Klein Center.

Email: jeffrey@metalab.harvard.edu

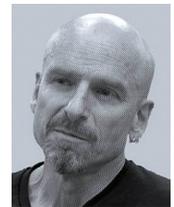

**Jeff** is the Director of Digital Infrastructure and Emerging Technology at the Harvard Art Museums. Areas of research include visualization of cultural datasets; open access to metadata and multimedia material; and data interoperability and sustainability. In November 2014, he helped launch the Lightbox Gallery, a public research and development space.

Email: jeff_steward@harvard.edu

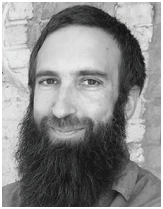

Rooted in the arts and humanities, straddling scholarly, critical, and creative practice, metaLAB is a community of scholars, designers, artists, makers, technologists, curators, and educators dedicated to modeling new forms of cultural communication, creative and critical practice, and knowledge production. Currently, metaLAB is based in Cambridge, MA, and Berlin.

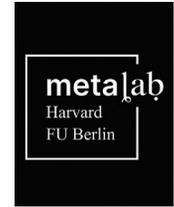